\documentclass[]{IEEEtran}
\usepackage{tmi}
\usepackage{cite}
\usepackage{amsmath,amssymb,amsfonts}
\usepackage{graphicx}
\usepackage{textcomp}
\usepackage{booktabs}   
\usepackage{subcaption}
\usepackage{makecell}
\usepackage[dvipsnames]{xcolor}
\usepackage{arydshln}
\usepackage{algorithm}
\usepackage{algpseudocode}
\usepackage{hyperref}       
\usepackage{url}            
\usepackage{nicefrac, xfrac}

\usepackage{nicefrac}
\usepackage{xcolor}

\usepackage{comment}

\newcommand{\mtrx}[1]{\mathbf{#1}}

\newcommand{\mytimes}{{\mkern-2mu\times\mkern-2mu}}

\def\BibTeX{{\rm B\kern-.05em{\sc i\kern-.025em b}\kern-.08em
    T\kern-.1667em\lower.7ex\hbox{E}\kern-.125emX}}
\begin{document}
\bstctlcite{IEEEexample:BSTcontrol}
\title{Morph-SSL: Self-Supervision with Longitudinal Morphing to Predict AMD Progression from OCT}
\author{Arunava Chakravarty, 
Taha Emre, Oliver Leingang, Sophie Riedl, Julia Mai, Hendrik P. N. Scholl, Sobha Sivaprasad, Daniel Rueckert, Andrew Lotery, Ursula Schmidt-Erfurth, and Hrvoje Bogunovi\'c
\thanks{Manuscript submitted March, 30, 2023. This work is supported in part by a Wellcome Trust Collaborative Award (PINNACLE) Ref. 210572/Z/18/Z, and FWF (Austrian Science Fund; grant number FG 9-N).}  
\thanks{A. Chakravarty, T. Emre, O. Leingang, S. Riedl, J. Mai, U. Schmidt-Erfurth, and H. Bogunovi\'{c} are with the Department of Ophthalmology and Optometry, Medical University of Vienna, Austria. H. Bogunovi\'{c} is also with the
Christian Doppler Lab for Artificial Intelligence in Retina, Medical University of Vienna, Austria (hrvoje.bogunovic@meduniwien.ac.at).} 
\thanks{D. Rueckert is with BioMedIA, Imperial College London, United
Kingdom, and also with the Institute for AI and Informatics in Medicine, Klinikum rechts der Isar, Technical University
Munich, Germany.}
\thanks{S. Sivaprasad is with the NIHR Moorfields Biomedical Research Centre,
Moorfields Eye Hospital NHS Foundation Trust, London, United Kingdom.}
\thanks{H. Scholl is with Institute of Molecular and Clinical Ophthalmology Basel, Switzerland, and also with the Department of Ophthalmology, University of Basel, Switzerland}
\thanks{A. Lotery is with Clinical and Experimental Sciences, Faculty of Medicine, University of
Southampton, United Kingdom}
}

\maketitle
\begin{abstract}
The lack of reliable biomarkers makes predicting the conversion from intermediate to neovascular age-related macular degeneration (iAMD, nAMD) a challenging task.  We develop a Deep Learning (DL) model to predict the future risk of conversion of an eye from iAMD to nAMD from its current OCT scan. Although eye clinics generate vast amounts of longitudinal OCT scans to monitor AMD progression, only a small subset can be manually labeled for supervised DL. To address this issue, we propose Morph-SSL, a novel Self-supervised Learning (SSL) method for longitudinal data. It uses pairs of unlabelled OCT scans from different visits and involves morphing the scan from the previous visit to the next. The Decoder predicts the transformation for morphing and ensures a smooth feature manifold that can generate intermediate scans between visits through linear interpolation. Next, the Morph-SSL trained features are input to a Classifier which is trained in a supervised manner to model the cumulative probability distribution of the time to conversion with a sigmoidal function. Morph-SSL was trained on unlabelled scans of 399 eyes (3570 visits). The Classifier was evaluated with a five-fold cross-validation on 2418 scans from 343 eyes with clinical labels of the conversion date. The Morph-SSL features achieved an AUC of 0.766 in predicting the conversion to nAMD within the next 6 months, outperforming the same network when trained end-to-end from scratch or pre-trained with popular SSL methods.  Automated prediction of the future risk of nAMD onset can enable timely treatment and individualized AMD management. 
\end{abstract}


\section{Introduction}
\label{sec:Introduction}

Age-related macular degeneration (AMD) is a leading cause of blindness in the elderly population\cite{wong2014global}. Although asymptomatic in its early and intermediate stages, it gradually progresses to a  late stage leading to irreversible vision loss.The early or intermediate AMD (iAMD) is primarily  characterized by the presence of drusen. Additionally, the Retinal Pigment Epithelium (RPE) and  Photoreceptor (PR) layers  degenerate over time and are associated with Hyper-reflective Foci (HRF).  The late stage is characterized by significant vision loss either due to the presence of Geographic Atrophy (GA) called dry AMD, the presence of choroidal neovascularisation (CNV) called neovascular AMD (nAMD), or a combination of both. nAMD is caused by the abnormal growth of blood vessels that leak fluid into the retina\cite{hallak2019imaging} which can be effectively treated with intravitreal anti-VEGF injections. If patients at a higher risk of conversion to nAMD can be identified in the iAMD stage itself, then potential future vision loss could be avoided through frequent monitoring and early treatment. However, the rate of progression varies widely across patients. There are no reliable biomarkers in the iAMD stage to differentiate between slow and fast progressors making it difficult for clinicians to determine the precise risk and timing of conversion. Thus, deep learning (DL) based methods to predict the future risk of conversion to nAMD can play a critical role in enabling patient-specific disease management. 

Optical Coherence Tomography (OCT) provides a 3D view of the retinal tissue and comprises a series of cross-sectional 2D image slices called B-scans. 
In clinical practice, a longitudinal series of OCT scans is routinely acquired over multiple patient visits to assess and monitor AMD progression. It generates a large amount of retrospective imaging data that can potentially be used to train DL models. However, due to the time, effort, and clinical expertise required, manual Ground Truth (GT) labels are rarely available for supervised training. Self-Supervised Learning (SSL) offers a way to address this issue by training DL networks to solve \textit{pretext} tasks on unlabelled training data to learn useful feature representations.

In this work, we propose a novel SSL method specifically adapted to longitudinal datasets called Morph-SSL. It involves morphing an OCT scan from one visit to a future visit scan of the same eye. We surmise that the change between the features extracted from two visits should reflect the structural deformation and the intensity changes between them. 
Morph-SSL is employed to develop a prognostic model to predict the future conversion from iAMD to nAMD within the next $t$ months from a single current OCT scan. $t$ can be any continuous time-point up to a maximum of 18 months. We refer to this task as TTC, predicting the probability distribution of the \textit{Time-to-Conversion}. Once an Encoder has been trained with Morph-SSL, a 3-layer Classifier is trained for the TTC task on limited labelled data. The Encoder and Classifier can further be fine-tuned jointly. Our key contributions are: 


(i) We propose Morph-SSL to learn features that capture temporal changes in the retinal tissue. It requires at least two visits per eye and can also be trained on scans acquired at irregular intervals. The learned features are semantically meaningful and can generate intermediate scans through linear interpolation with smooth transition between two visits.

(ii) We model the Cumulative Distribution Function (CDF) of the probability of the time to conversion with a sigmoidal function over time. It allows using continuous GT labels of conversion time during training, ensures the monotonic non-decreasing property of the CDF, and can predict the conversion risk for arbitrary continuous time-points at test time.

(iii) We propose a score $r \in [0,1]$ that quantifies the future risk of OCT scans to convert to nAMD and can categorize them into low, moderate, and high risk groups.

(iv) We develop an efficient CNN network to process entire OCT volumes instead of individual 2D B-scans. We explore (a) \textit{S3DConv} block to replace 3D convolutions with three groups of 2D convolutions oriented in the three orthogonal planes; (b) concatenation-based (instead of additive) skip connections to have same output channel size with fewer convolutions; (c) Layer Normalization instead of Batch Normalization to allow training with a batch size of 1.

\section{Related Work}
\label{sec:Related_Work}
\textbf{Self-Supervised Learning:} It offers a way to overcome the paucity of labelled datasets for supervised training. SSL learns feature representations from unlabelled data by training the network on a pretext task that does not need manual labels. The SSL-trained models can either be utilized for off-the-shelf feature extraction or to provide initial weights for fine-tuning on the desired \textit{downstream} task with limited labelled training data. The recent SSL methods employ pretext tasks based on image reconstruction or Contrastive Learning (CL). The reconstruction based methods train networks to predict the original image from its distorted version and have been applied to X-ray, CT, MRI and ultrasound images  \cite{model_genesis},\cite{chen2019self}. The distortions involve transformations such as non-linear intensity mapping, local shuffling and in-painting in Model-Genesis\cite{model_genesis} and randomly swapping patches in the image \cite{chen2019self}.

CL has been applied to chest X-ray, dermatology \cite{azizi2021big}, histology \cite{ciga2022self}, MRI \cite{zeng2021positional} and ultrasound \cite{chen2021uscl} images. CL trains networks using random batches comprising two data-augmented versions per image, called \textit{positive pairs}. While positive pairs are pulled closer, the features of different images in the batch called  \textit{negative pairs} are pushed apart. However, the images in a negative pair can still be semantically similar (same pathology or disease stage), resulting in many \textit{False Negative pairs}. Their impact can be reduced by training with large batch sizes (1024 for chest X-rays, 512 for dermatology images in \cite{azizi2021big} and 128 for histology image patches in \cite{ciga2022self} ). Since large batch sizes do not scale well to 3D images due to limited GPU memory, existing methods learn features at a 2D, slice-level for 3D MRI volumes \cite{zeng2021positional}, or for individual frames in ultrasound videos \cite{chen2021uscl} where neighboring slices/frames of the same 3D image are excluded from negative pairs. The recently proposed \textit{Non-Contrastive} methods overcome the problem of \textit{False Negative pairs}. They do not maximize the negative pair separation but only ensure that they do not collapse onto the same feature representation. VICReg \cite{bardes2021vicreg} keeps the standard deviation of each feature dimension over a batch above a threshold. Barlow Twin \cite{zbontar2021barlow} forces the cross-correlation between two batch of features extracted from the two images in each positive pair to be close to the identity matrix. BYOL\cite{grill2020bootstrap} prevents feature collapse using slightly different network weights to extract features for the two views in the positive pair, where the second network weight is computed as the moving average of the past weights.

CL and Non-Contrastive SSL have been adapted for retinal OCT to learn features for 2D B-scans with training batch sizes of 128 in \cite{holland2022metadata} and 384 in \cite{emre2022tinc}. Another method learns features for central B-scans by predicting the time interval between two input scans from random visits of the same patient \cite{rivail2019modeling}. In contrast, \textit{Morph-SSL with a novel image morphing based pre-text task can be trained with a batch size of 1 to reduce GPU memory usage, allowing us to learn feature representations for entire 3D OCT volumes instead of 2D B-scans.}


\begin{figure*}[!t]
 \centering
  \includegraphics[width=.8\textwidth]{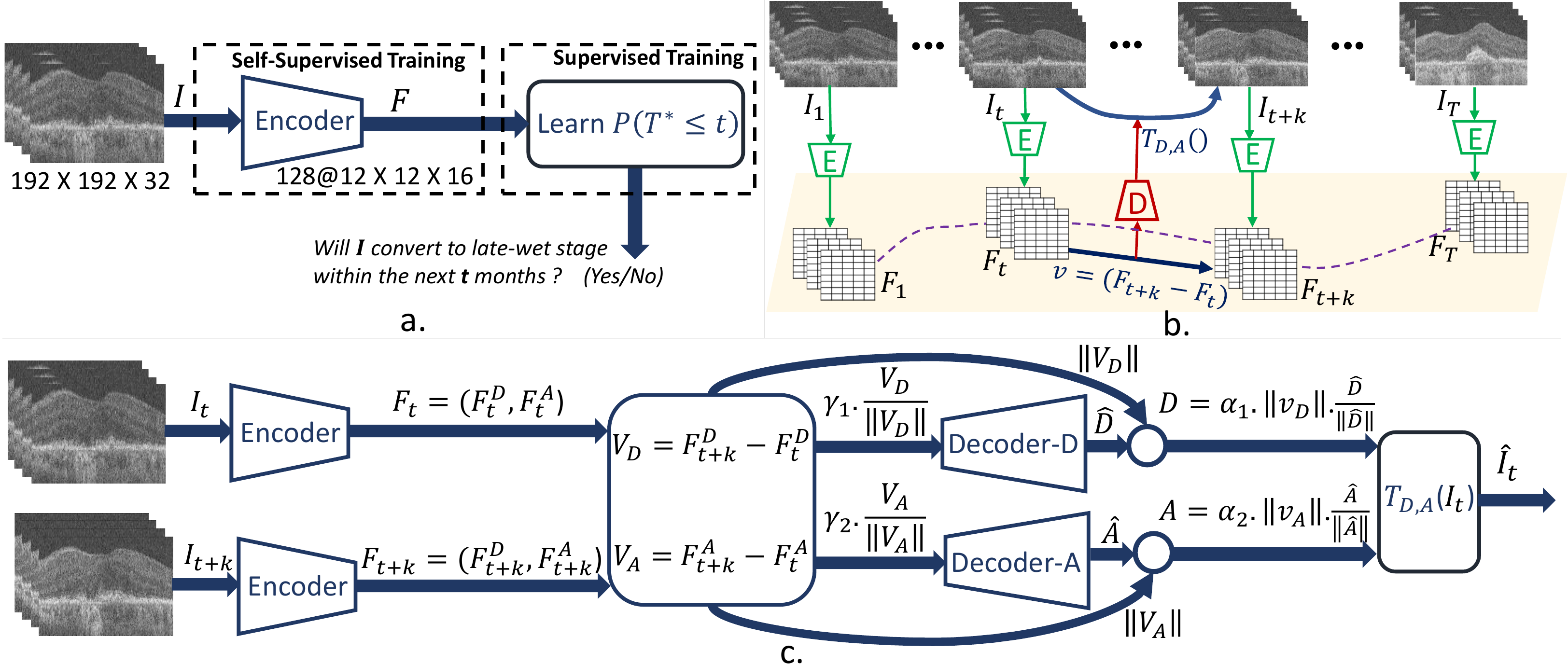}
\caption{\small (a) An overview of our 2-stage training framework. The motivation behind Morph-SSL is shown in (b) and its details in (c).}
\label{fig:intro} 
 \end{figure*}
\textbf{Time to Conversion Prediction:} 
Existing methods either employ Color Fundus Photographs (CFP) or OCT imaging for TTC prediction. CFP is a 2D image of the retinal surface and lacks a cross-sectional view of the retina. A 9-grade AREDS disease severity scale \cite{davis2005age} further stratifies the iAMD stage in CFPs, where each successive stage has been linked to an increased 5-year risk of conversion to advanced AMD  (from  $1\%$ in grade 1 to about $50\%$ risk in grade 9). However, no such severity scale exists for the relatively new OCT imaging.

Some \textit{CFP-based methods} predict the AREDS severity scale \cite{burlina2018use}, \cite{bhuiyan2020artificial}. Two-year conversion of nAMD was predicted with an ensemble of such predictions combined with features from drusen segmentation and demographic data \cite{bhuiyan2020artificial}. The CNN-LSTM based methods in \cite{bridge2020development}, \cite{yin2022predicting} require images from multiple past visits, hence cannot be used for patients visiting for the first time. The input CFPs from visits at irregular intervals are handled by scaling the input image features with visit time intervals \cite{bridge2020development} or using a time-aware LSTM network\cite{yin2022predicting}. In \cite{Ganjdanesh2022longl}, a Generative Adversarial Network was used to generate synthetic CFP images for future time-points. Combining CFP with genetic features can improve performance \cite{yan2020deep}, but such information is not readily available in eye clinics. While CFP-based methods can predict long-term conversion, they are not sensitive to short-term conversion risks within 2 years, required for effective clinical intervention. Because CFPs lack a 3D view of the retina, they cannot capture subtle changes in retinal layers or extract accurate lesion volumes.

Many \textit{OCT-based methods} first extract a set of handcrafted quantitative biomarkers to capture the distribution, appearance and volume of lesions like drusen, HRF and retinal layers such as RPE and PR. These biomarkers combined with other demographic \cite{banerjee2020prediction} or genetic data \cite{schmidt2018prediction} is input to an LSTM \cite{banerjee2020prediction}, Cox proportional hazards model \cite{schmidt2018prediction}, or an L1-penalized Poisson model \cite{de2014quantitative} to predict the TTC. The biomarkers are extracted with automated segmentation methods that are often inaccurate and require voxel-level labels to train. Moreover, handcrafted biomarkers may not adequately capture the subtle retinal changes related to disease progression.
 
Another approach directly uses the OCT scans as input. To reduce the compute and GPU memory, most existing methods operate on individual B-scans with 2D CNNs. Some methods in \cite{rivail2019modeling},  \cite{holland2022metadata} only use the central B-scan that passes through the macula, ignoring the remaining B-scans in the volume. While \cite{rivail2019modeling} employed this strategy for TTC prediction, \cite{holland2022metadata} used it on other tasks such as predicting age, sex and visual acuity from OCT. Alternatively, 2D CNNs can be applied independently to every B-scan in the volume. During inference, the predictions from each B-scan is pooled, 
either by taking the the average \cite{russakoff2019deep} or maximum \cite{emre2022tinc} to obtain the volume-level prediction. During training, the same GT for the conversion time is used for every B-scan in the volume, even if only a few of them have the  biomarkers indicative of progression risk, resulting in noisy training labels. \textit{In contrast to these methods, in this work we explore a full 3D approach by developing a compact 3D-CNN network to effectively capture the spatial information across the individual B-scans}. Other than our work, the only other 3D-CNN method is found in \cite{yim2020predicting}, which employs two prediction networks to predict the conversion risk within six months, one using raw OCT volumes and the other using retinal layer and lesion segmentation maps.

\section{Method}
\label{sec:Method}
The proposed method depicted in Fig. \ref{fig:intro} a. is trained in two stages. First, a Fully Convolutional Encoder is trained with Morph-SSL to leverage unlabelled data (see Section \ref{sub_section:SSL}). It projects an input OCT scan $\mtrx{I}$ to a convolutional feature map $\mtrx{F}$. Next, in Section \ref{sub_section:Classifier}, the CDF of the TTC is modeled as a sigmoidal function over time. Its parameters are predicted by a classifier trained in a supervised manner with $\mtrx{F}$ as input.

\subsection{Self-Supervised Learning}
\label{sub_section:SSL}

\textbf{Motivation:} Let $\lbrace \mtrx{I}_t | 1\le t \le T \rbrace$ represent a set of 3D OCT scans of an eye acquired over $T$ visits, in which $\mtrx{I}_t$ is taken on the $t^{th}$ visit. The Encoder projects each $\mtrx{I_t}$ to a feature map $\mtrx{F}_t$ of size $128@12\mytimes12\mytimes16$. $\mtrx{F}_t$ can be interpreted as 128-dimensional features for overlapping 3D image patches in $\mtrx{I}_t$, with the patch size defined by the effective receptive field of the Encoder. As AMD progresses over successive visits, $\mtrx{F}_t$ traces a trajectory (denoted by the violet dotted line in Fig \ref{fig:intro} b.) that is locally linear between nearby visits $\mtrx{I}_t$, and $\mtrx{I}_{t+k}$ (assuming a smooth feature manifold) but maybe non-linear over the entire AMD progression. Let $\mathcal{T}_{D,A}(.)$ denote a transformation which morphs $\mtrx{I}_t$ to look similar to $\mtrx{I}_{t+k}$, with parameters $\mtrx{D}$ and $\mtrx{A}$. As $\mtrx{I}_t$ morphs into $\mtrx{I}_{t+k}$ in the image space, $\mtrx{F}_t$ should be linearly displaced to $\mtrx{F}_{t+k}$ by $\mtrx{V}_{t}=\mtrx{F}_{t+k}-\mtrx{F}_{t}$ in the feature manifold. This observation motivates our pretext task for Morph-SSL which employs an Encoder-Decoder architecture. The Encoder projects scans from two nearby visits, $\mtrx{I}_t$ and $\mtrx{I}_{t+k}$ to their features $\mtrx{F}_t$ and $\mtrx{F}_{t+k}$. The Decoder uses the displacement $\mtrx{V}_{t}$ as input to predict $\mtrx{D}$ and $\mtrx{A}$ of the morphing transformation $\mathcal{T}_{D,A}$. Our pretext task ensures that the displacements in the learned feature manifold capture the corresponding appearance changes in the image space.  

$\mathcal{T}_{D,A}$ comprises a spatial deformation with the 3-channel $\mtrx{D}$ and an additive intensity transformation with the 1-channel $\mtrx{A}$, both of the same spatial size as $\mtrx{I}_t$. Each voxel at location $\Vec{p}$ in $\mtrx{I}_t$ is displaced to the location $\Vec{p}+\mtrx{D}(\Vec{p})$, where $\mtrx{D}(\Vec{p})$ is a 3-dimensional displacement vector. Additionally, $\mtrx{A}(\Vec{p})$ captures the intensity changes at each location $\Vec{p}$, caused by newly formed pathologies in $\mtrx{I}_{t+k}$ such as fluids or drusen. Thus, the transformed image $\mtrx{\widehat{I}}_{t}=\mathcal{T}_{D,A}\left( \mtrx{I}_{t}\right)=\Phi\left( \mtrx{I}_t; \mtrx{D}\right)+\mtrx{A}$, where $\Phi$ is the spatial deformation applied in a differentiable manner similar to the registration methods in \cite{balakrishnan2019voxelmorph}, \cite{zhang2018inverse} based on the Spatial Transformer Networks \cite{jaderberg2015spatial}. 

\textbf{Morph-SSL framework:} The details are depicted in Fig.\ref{fig:intro}c. $\mtrx{F_t}$ is split into two subspaces $\mtrx{F}^{D}_{t}$ and $\mtrx{F}^{A}_{t}$ of 64 channels each. The Decoder has two sub-networks, \textit{Decoder-D} and \textit{Decoder-A} that operate on $\mtrx{F}^{D}_{t}$ and $\mtrx{F}^{A}_{t}$ feature maps respectively, to predict $\mtrx{D}$ and $\mtrx{A}$. A notion of semantically meaningful directions and distance is incorporated. The amount of deformation between $\mtrx{I}_t$ and $\mtrx{I}_{t+k}$ should be proportional to the Euclidean distance $||\Vec{V}_{D} ||=||\Vec{F}^{D}_{t+k}-\Vec{F}^{D}_{t}||_2$, while the nature and location of the deformation should be captured by the direction alone, represented by the unit vector $\Vec{V}_{D}/||\Vec{V}_{D}||$. This property is enforced by our Decoder architecture. Only the direction information $\gamma_1.(\Vec{V}_{D}/||\Vec{V}_{D}||)$ is input to \textit{Decoder-D} and its output $\mtrx{\widehat{D}}$ is normalized and scaled to obtain $\mtrx{D}=\alpha_1. ||\Vec{V}_D||.(\mtrx{\widehat{D}}/||\mtrx{\widehat{D}}||)$. This ensures that $||\mtrx{D}||=\alpha_1.||\mtrx{V}_D||$. Both $\gamma_1$, $\alpha_1$ are learnable parameters (positive scalar weights) employed for numerical stability during training. A similar scheme is employed to predict $\mtrx{A}$. The direction $\gamma_2. (\Vec{V}_{A}/|| \Vec{V}_{A} ||)$ is input to \textit{Decoder-A} and its output scaled to $\mtrx{A}=\alpha_2. ||\Vec{V}_A||.(\mtrx{\widehat{A}}/|| \mtrx{\widehat{A}}||)$, where $\gamma_2$,  $\alpha_2$ are learnable positive weights (see Fig. \ref{fig:intro} c).
\begin{figure*}[!t]
 \centering
  \includegraphics[width=0.90\textwidth]{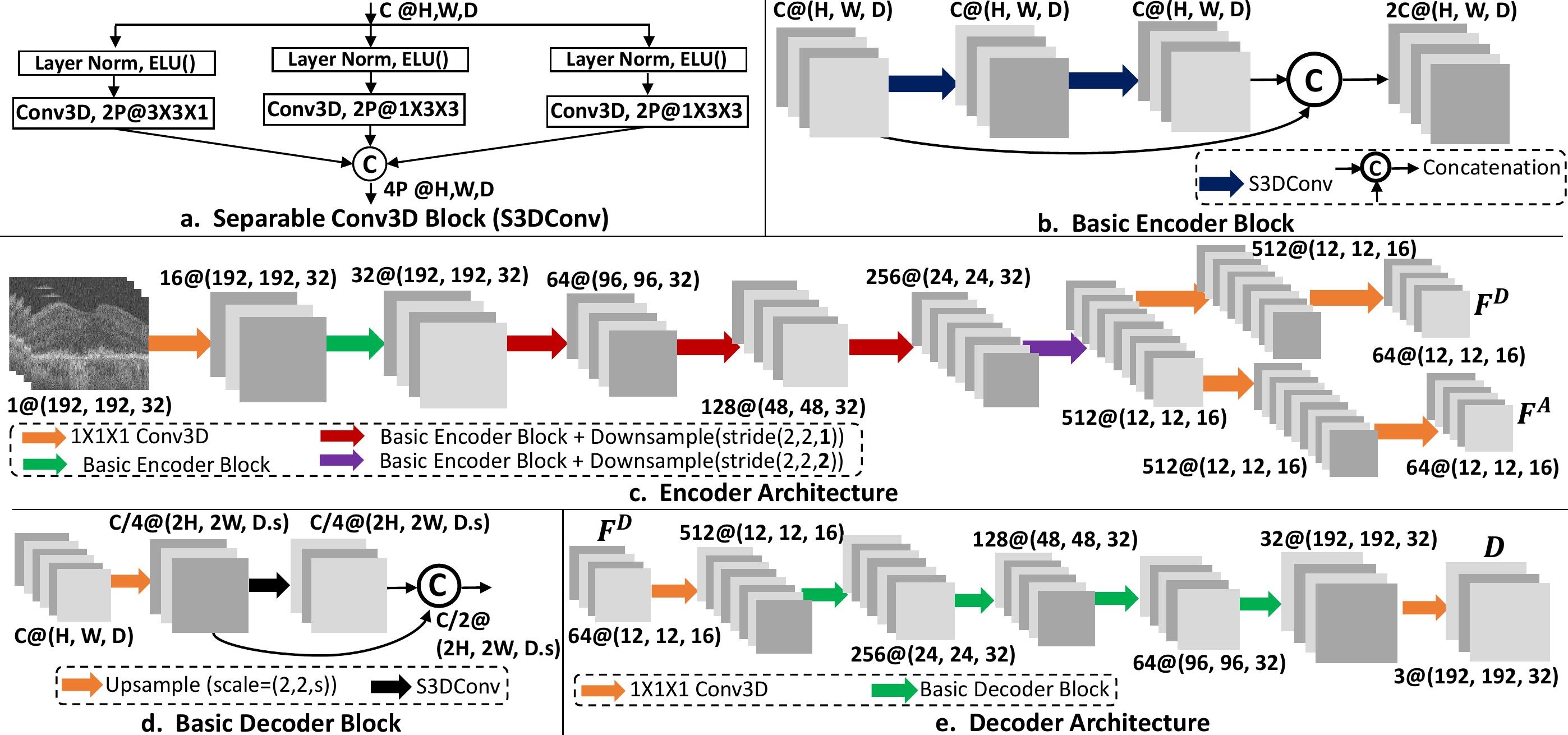}
\caption{\small Our Encoder (c) comprises a series of Basic Encoder Blocks (b). Except for the number of output channels in the last layer, both Decoder-D and Decoder-A have the same architecture (e) and consist of a series of Basic Decoder Blocks (d). S3DConv (a) is used as the basic convolution operation in both the Basic Encoder and Decoder Blocks.}
\label{fig:arch} 
 \end{figure*}
 
\textbf{Loss Function:} The Encoder-Decoder network is trained to minimize the Mean Squared Error (MSE) between $\mtrx{\widehat{I}}_{t}$ and $\mtrx{I}_{t+k}$ by comparing their voxel intensities ($\mathcal{L}_{mse}$) as well as their feature maps extracted with a CNN ($\mathcal{L}_{prc}$). $\mathcal{L}_{mse}$ alone leads to blurred reconstructions which is remedied by using the additional \textit{perceptual loss} $\mathcal{L}_{prc}$  \cite{dosovitskiy2016generating},\cite{yang2018low}. The background noisy region is ignored while computing $\mathcal{L}_{mse}$ and $\mathcal{L}_{prc}$ using a binary Region of Interest (ROI) mask of the retinal tissue, $\mtrx{R}_t$ (for $\mtrx{I}_t$) and $\mtrx{R}_{t+k}$ (for $\mtrx{I}_{t+k}$), obtained during pre-processing. Before computing the loss, the background regions are masked out through element-wise multiplication $\mtrx{I}_{t+k}=\mtrx{I}_{t+k}\odot \mtrx{R}_{t+k}$  and $\mtrx{\widehat{I}}_{t}=\mtrx{\widehat{I}}_{t}\odot \Phi(\mtrx{R}_{t}; \mtrx{D})$. The Encoder does not require the binary masks at inference time as it is only used to compute the loss.

$\mathcal{L}_{mse}$ has two terms. First, the MSE is computed with the only spatially deformed image $\Phi(\mtrx{I}_{t}; \mtrx{D})$. Next, $\mtrx{A}$ is fitted to the residual difference left after the spatial deformation, $\mtrx{U}=\mtrx{I}_{t+k}-\Phi(\mtrx{I}_{t};\mtrx{D}).detach()$.  The $detach()$ indicates that the gradients are not allowed to backpropagate through $\mtrx{U}$ which is computed on the fly and treated as the GT for $\mtrx{A}$. This two step design is to ensure that $\mtrx{D}$ accounts for most of the reconstruction and avoid trivial solutions where $\mtrx{D}$ is an identity transformation (0 displacement for all voxels) while $\mtrx{A}$ tries to learn the entire difference $\mtrx{I}_{t+k}-\mtrx{I}_{t}$. Thus, 
\begin{equation}
\mathcal{L}_{mse}=\frac{\lambda_1}{|\Omega|}.|| \mtrx{I}_{t+k}- \Phi(\mtrx{I}_{t}; \mtrx{D})||_2^2 + \frac{\lambda_2}{|\Omega|}.|| \mtrx{U}-\mtrx{A}||_2^2,
\end{equation}
where $|\Omega|$ is the total number of voxels in the image and the relative weights $\lambda_1=10^1$, $\lambda_2=10^2$ were set empirically.

Instead of using a standard pre-trained CNN, we copy the first 3 layers of our Encoder itself to build a Comparator Network $\psi$ which is then used for the perceptual loss, 
\begin{equation}
\mathcal{L}_{prc}=\frac{1}{3}\sum _{j=1} ^ 3 \frac{1}{|\Omega|} || \psi_{j}(\mtrx{I}_{t+k})- \psi_{j}(\mtrx{\widehat{I}}_t)||_2^2.
\end{equation}
$\psi_j$ denotes the output of the $j^{th}$ layer of $\psi$. The network weights of $\psi$ are not updated through backpropagation as it may learn to collapse the features of $\mtrx{I}_{t+k}$ and $\mtrx{\widehat{I}}_t$, even though they appear different. Inspired by BYOL \cite{grill2020bootstrap}, we update $\psi$ with an exponential moving average of the Encoder weights obtained while it is being trained.

Additional \textit{regularization} loss terms are also incorporated to obtain an anatomically feasible $\mathcal{T}_{D,A}$. $\mtrx{D}$ is encouraged to be diffeomorphic by penalizing it to be smooth with $\mathcal{L}_{smt}$ and prevent folding with $\mathcal{L}_{fld}$. The $\mathcal{L}_{smth}=\sum _{p \in \Omega} || \nabla \mtrx{D}(\Vec{p})||^2_2$ was defined as in \cite{balakrishnan2019voxelmorph}, where  the spatial gradient $\nabla \mtrx{D}(\mtrx{p})$ is computed at all voxel positions through discrete numerical approximation. $\mathcal{L}_{fld}$ as defined in \cite{zhang2018inverse}, penalizes the anatomically infeasible deformations where the retinal tissue folds onto itself.
Finally, the sparsity of $\mtrx{A}$ is ensured with an L1-regularization $\mathcal{L}_{add}=\sum _{p \in \Omega} | \mtrx{A}(\Vec{p})|$. Thus the total loss is
\begin{equation}
\mathcal{L}=\mathcal{L}_{mse}+\lambda_3.\mathcal{L}_{prc}+\lambda_4.\mathcal{L}_{smt}+\lambda_5.\mathcal{L}_{fld}+\lambda_6\mathcal{L}_{add},
\end{equation}
where $\lambda_3=10^{1}$, $\lambda_4=10^{-1}$, $\lambda_5=10^{6}$  and $\lambda_6=10^{-5}$ are empirically fixed, based on their relative importance and also to scale the different loss terms to a similar range. The range of the $\mathcal{L}_{fld}$ is orders of magnitude lower than the other terms, thus requiring a significantly larger scaling weight.

\textbf{Network Architecture:} The \textit{separable 3D Convolution Block} (S3DConv) depicted in Fig. \ref{fig:arch}a. replaces 3D convolutions throughout our Encoder and Decoder Networks to reduce computation and network parameters. It employs 2D convolution filters in the three orthogonal planes. While $50 \%$ of the filters are $3\mytimes 3\mytimes 1$ that operate on individual B-scans, the remaining are an equal number of $1 \mytimes 3 \mytimes 3$ and $3 \mytimes 1 \mytimes 3$ filters to capture contextual information across the neighboring B-scans. Using Layer Normalization instead of Batch Normalization allows training with a batch size of 1. The \textit{pre-activation} strategy \cite{he2016identity} ensures that the normalization and $ELU$ activations are applied after the skip connections, at the beginning of the next S3DConv block for better gradient backpropagation.

The \textit{Encoder} depicted in Fig. \ref{fig:arch} c. has a series of five Basic Encoder Blocks interleaved with downsampling. The Basic Encoder Block comprises two S3DConv Blocks followed by a concatenation based skip connection (see Fig. \ref{fig:arch} b). Here, each S3DConv has $C$ input and output channels by setting $P=C/4$ in  Fig. \ref{fig:arch} a. The downsampling is performed with a strided $3\mytimes3\mytimes3$ depthwise-separable convolution \cite{chollet2017xception}. It applies a separate $3\mytimes3\mytimes3$ convolution (with 1 input and output channel) to each of the $C$ input channels individually and their outputs are concatenated together. It is implemented in Pytorch by setting \textit{groups}=1 in the Conv3D layer. Due to large voxel spacing across the B-scans, downsampling along this direction is only performed in the final block (violet arrow in Fig. \ref{fig:arch} c ) with a stride of $(2,2,2)$ to ensure a roughly isotropic receptive field. All previous downsampling layers (red arrows) use a $(2,2,1)$ stride to only halve the height and width dimensions. The last Encoder block is followed by two parallel pathways, each consisting of two $1\mytimes1\mytimes1$, 3D convolutional layers (orange arrows) to obtain the final 64 channel $\mtrx{F}^{D}$ and $\mtrx{F}^{A}$.

\textit{Decoder: }
Both \textit{Decoder-D} and \textit{Decoder-A} (in Fig. \ref{fig:intro} c.) have the same architecture as shown in Fig. \ref{fig:arch} e., except for the number of output channels in the last $1\mytimes 1 \mytimes 1$, convolution layer (orange arrow) to have a 1 (or 3) channel output for $\mtrx{A}$ (or $\mtrx{D}$). The Decoder architecture employs a series of Basic Decoder Blocks (green arrows). They map a $C@(H,W,D)$ input feature map to a  $\frac{C}{2}@(2H,2W,D.s)$ output, where $s$ is the upsampling factor across B-scans ($s=2$ in the first block, $1$ otherwise). As depicted in  Fig. \ref{fig:arch} d, it comprises an upsampling layer followed by a S3DConv whose outputs are concatenated with a skip connection. The upsampling layer performs two operations. First the input of size $C@(H,W,D)$ is upsampled to $C@(2H,2W,D.s)$ using trilinear interpolation. Next, a Depth-separable $3\mytimes3\mytimes3$ convolution is employed, which divides the $C$ input channels into $\frac{C}{4}$ groups of $4$ channels each. A separate convolution filter is applied to each group to compress them to a  single channel resulting in a $\frac{C}{4}@(2H,2W,D.s)$ output.
\subsection{Downstream TTC estimation Task}
\label{sub_section:Classifier}

\begin{figure*}[h]
 \centering
  \includegraphics[width=1.0\textwidth]{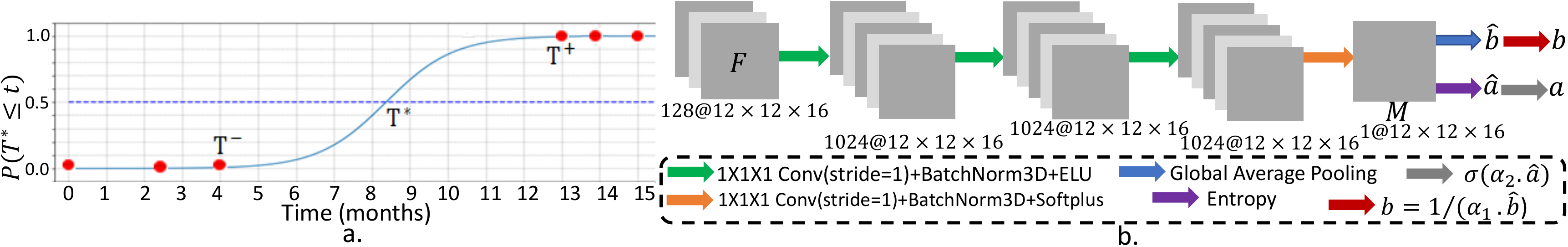}
\caption{\small Overview of the TTC Task. (a) CDF of the conversion time $T^*$ can be best modeled using a sigmoidal function. Exact $T^*$ is unknown due to the discrete nature of the visits (red dots) but occurs between the first visit where the eye has converted ($T^+$) and the visit ($T^-$) just before it. (b) The Classifier Network to predict the sigmoidal function parameters.}
\label{fig:classification} 
 \end{figure*}

The problem setting of the Downstream TTC task for an eye is depicted in Fig. \ref{fig:classification} a. An OCT is acquired at each visit (red dots) occuring at irregular time intervals. The eye remains in the early/iAMD stage up to the visit at time $T^-$ and is first diagnosed to have progressed to nAMD at time $T^+$. The exact time of conversion $ T^*$ is unknown as patients are monitored at discrete time-points but lies in $T^- < T^* \le T^+$. We treat $T^*$ as a continuous random variable and aim to model its CDF, $P(T^* \le t)$ (y-axis in Fig. \ref{fig:classification} a). $P(T^* \le t)$ is the probability that the eye has converted within the time-point $t$ . The binary GT for  $P(T^* \le t)$ is $0$ for $0\le t\le T^-$, 1 for $t\ge T^+$ and unknown in the range $T^{-}<t<T^+$. We propose to model $P(T^* \le t)$ with a sigmoidal distribution over time as

\begin{equation}
\label{eqn:cdf}
p_t=P(T^* \le t)=1 /\left [1+exp \left\lbrace - \left ( \frac{t-b}{a+0.05}\right ) \right\rbrace \right ],
\end{equation}
where $b$ is an estimate of $T^*$ and $a$ controls the slope of the sigmoidal CDF. A steep slope (small $a$) would indicate a fast progression rate around $T^*$ and viceversa.

\textbf{Classifier Architecture:} The scalar $a$ and $b$ are predicted with the classifier in Fig. \ref{fig:classification}b. The SSL-trained feature map $\mtrx{F}$ of the input OCT scan is fed to the classifier. $\mtrx{F}$ is mapped to a single channel feature map $\mtrx{M}$ through a series of three $1\mytimes1\mytimes1$ convolutional layers. A Class Activation Map (CAM) can be computed as a weighted sum of all channels in the final convolutional feature map, which in our case is $\mtrx{M}$ with a single channel. Thus, $\mtrx{M}$ can be interpreted as a saliency map for our classifier (see Fig. \ref{fig:sal})  which motivates how $a$ and $b$ is computed.

The $b$ is obtained through the Global Average Pooling (GAP) of $\mtrx{M}$ denoted by $\widehat{b}$, scaling it by a non-negative learnable scalar weight $\alpha _1$ and taking the reciprocal $b=1 / (\alpha _1 \cdot \widehat{b})$. We hypothesize that images predicted to convert soon (with small $b$) should lead to higher activations in the saliency map $\mtrx{M}$. 

The $a$ is obtained by computing the spatial entropy of $\mtrx{M}$ denoted by $\widehat{a}$, scaling it by non-negative learnable scalar $\alpha _2$ and applying the sigmoid activation. We hypothesize that low entropy (certain locations in $\mtrx{M}$ have high activations while others take very small values) indicates the detection of some salient regions in the OCT which may correlate to a sudden disease progression around the conversion event leading to a steep slope (small $a$). The spatial entropy is computed by first normalizing M to sum to 1, $\mtrx{M^{'}}(i)=\mtrx{M}(i) / \sum _{p \in \Omega} \mtrx{M}(p)$ and then computing the entropy as $H=-\sum _{i \in \Omega} \mtrx{M^{'}}(i). log \mtrx{M}^{'}(i)$, where $\Omega$ represents each spatial position in $\mtrx{M}$.

\textbf{Loss Function:} 
A maximum time interval of 18 months (normalized to [0,1]) was considered as longer durations are not useful for clinical intervention. The $T^*$ for scans that do not convert within 18 months is unknown. For each scan, the classification loss $\mathcal{L}_{cls}$ consists of the average binary cross entropy loss (BCE) computed for two time-points as
\begin{equation}
\mathcal{L}_{cls} = \begin{cases}
\mathcal{L}_{ce}\left(p_{T^+}, 1 \right) + \mathcal{L}_{ce}\left(p_{T^-}, 0\right) \text{, if $0 \le T^+, T^- \le 1$} \\
\mathcal{L}_{ce}\left(p_{0}, 0 \right) + \mathcal{L}_{ce}\left(p_{1}, 0\right) \text{, if $ T^+, T^- > 1$} \\
\mathcal{L}_{ce}\left(p_{0}, 1 \right) + \mathcal{L}_{ce}\left(p_{1}, 1\right) \text{, if $ T^+ = 0$}, \\
\end{cases}
\label{eqn_classification}
\end{equation}
where $p_t$ at time $t$ is computed using eq. \ref{eqn:cdf}. $\mathcal{L}_{ce}$ denotes half of BCE loss to compute the average at the two time-points. The first condition in eq. \ref{eqn_classification} occurs when both $T^-$ and $T^+$ occur within 18 months (1 after normalization) and $\mathcal{L}_{ce}$ is computed at these two time-points with the GT labels 0 at $T^-$ and 1 at $T^+$. Since sigmoidal function is monotonically non-decreasing, minimizing the loss at these two points automatically improves $p_t$ for all $t$ because $p_{T^-} \approx 0$ also ensures equal or lower predictions before $T^-$ and $p_{T^+} \approx 1$ enforces equal or higher predictions after $T^+$.  In the second condition in eq. \ref{eqn_classification}, the conversion (if the scan ever converts) occurs after 18 months and exact  $T^+$ and $T^-$ are unknown. Here, $\mathcal{L}_{ce}$ is computed at $t=0$ and $1$ with a GT label of 0 in both cases. The last condition in eq. \ref{eqn_classification} occurs when the input OCT scan is the first visit of conversion and the GT label remains 1 throughout the 18-month interval. In addition to $\mathcal{L}_{cls}$, two regularization terms are also employed. Thus, the total loss 

\begin{equation}
\mathcal{L}_{tot}=\mathcal{L}_{cls}+ \gamma_1 ||a||_2^2 + \gamma_2 || \mtrx{M}\odot ( 1-\mtrx{\widehat{R}} )||_1,
\label{eqn_ttc}
\end{equation}
where $\gamma_1=\gamma_2=0.1$. An L2-regularization of $a$ is performed for numerical stability. Moreover, higher activations in $\mtrx{M}$ outside the retina defined by the binary mask $\mtrx{\widehat{R}}$ are penalized. $\mtrx{\widehat{R}}$ is the  ROI mask of the input OCT resized to $12\mytimes12\mytimes16$.

\setlength{\tabcolsep}{5pt}
\renewcommand{\arraystretch}{1.2}
\begin{table*}[!t]
\centering
\caption{\small Area under the ROC curve(AUC) and Balanced Accuracy (mean$\pm$std. deviation)  across five-folds on the downstream task dataset. A DeLong test between the AUCs for each time-point between the Proposed-SSL-Freeze (row 2) vs the rest is performed and the values highlighted with $^*$ are \textbf{not} statistically different with $p>0.05$}
\resizebox{0.985\textwidth}{!}{
\begin{tabular}{llllllllll}
\toprule
SL.& & \multicolumn{2}{c}{0 month}         & \multicolumn{2}{c}{6 month}      & \multicolumn{2}{c}{12 month}    & \multicolumn{2}{c}{18 month}     \\
No. & & AUROC & Bal Acc. & AUROC & Bal Acc. & AUROC & Bal Acc. & AUROC & Bal Acc. \\ \midrule 

\multicolumn{10}{l}{Ablation on the impact of SSL} \\ 
\midrule
1 & Proposed-Random Init & $0.808\pm0.04$  & $0.724\pm0.05$  & $0.706\pm0.03$ & $0.624\pm0.05$  & $0.668\pm0.05$ & $0.567\pm0.04$ & $0.655\pm0.08$ & $0.539\pm0.04$ \\ 
2 & Proposed-SSL-Freeze & $0.881\pm0.02$ & $0.781\pm0.02$      & $0.766\pm0.02$ & $0.706\pm0.01$      & $0.714\pm0.04$ & $0.641\pm0.06$      & $0.687\pm0.06$ & $0.590\pm0.08$ \\

3 & Proposed-SSL-Finetune & $0.876\pm0.02^{*}$ & $0.805\pm0.03$  & $0.766\pm0.02^*$ & $0.729\pm0.04$  & $0.716\pm0.04^*$ & $0.666\pm0.09$  & $0.693\pm0.06$ & $0.620\pm0.12$ \\
\midrule 
\multicolumn{10}{l}{Ablation on Classification Loss in eq. \ref{eqn_classification}. SSL trained weights are frozen, only the Classifier is trained.}  \\ 
\midrule 
4 & no $||a||_2^2$  & $0.874\pm0.02^{*}$ & $0.797\pm0.02$ & $0.753\pm0.02$ & $0.695\pm0.01$ & $0.703\pm0.04$ & $0.640\pm0.04$ & $0.675\pm0.06$ & $0.605\pm0.06$ \\
5 & no $|| \lbrace\mtrx{M}\odot ( 1-\mtrx{\widehat{R}} )\rbrace _+||_1$ & $0.870\pm0.03$ & $0.779\pm0.04$ & $0.754\pm0.03$  & $0.697\pm0.03$ & $0.709\pm0.06^*$ & $0.612\pm0.07$ & $0.688\pm0.07^*$ & $0.566\pm0.06$ \\
\midrule
\multicolumn{10}{l}{Ablation on the Classifier Architecture. SSL trained weights are frozen, only the Classifier is trained.} \\ 
\midrule
6 & Multilabel Classifier & $0.903\pm0.01$ & $0.707\pm0.07$ & $0.756\pm0.02^*$ & $0.665\pm0.02$ & $0.705\pm0.05^*$ & $0.649\pm0.02$ & $0.676\pm0.07$ & $0.616\pm0.05$ \\ 
7 & Separate $a$ prediction & $0.873\pm0.03^*$ & $0.787\pm0.03$ & $0.765\pm0.02^*$ & $0.695\pm0.02$ & $0.715\pm0.04^*$ & $0.626\pm0.06$ & $0.684\pm0.06^*$ & $0.576\pm0.07$\\
\midrule 
\multicolumn{10}{l}{Benchmarking against standard 3D CNN networks. The entire network is fine-tuned after initalization with weights pre-trained on Kinetics dataset.} \\ 
\midrule
8 & I3D \cite{carreira2017quo} & $0.803\pm0.05$  & $0.743\pm0.04$  & $0.700\pm0.03$   &  $0.630\pm0.04$  & $0.655\pm0.03$ & $0.590\pm0.04$ & $0.648\pm0.02$ & $0.570\pm0.04$ \\
9 & X3D \cite{feichtenhofer2020x3d} & $0.797\pm0.02$  & $0.692\pm0.03$  & $0.711\pm0.01$  & $0.668\pm0.02$   & $0.673\pm0.02$ & $0.616\pm0.03$ & $0.668\pm0.02$ & $0.579\pm0.04$ \\
\bottomrule
\end{tabular}
}
\label{Tab:Result_Pred1}
\end{table*}

\section{Experiments and Results}
\textbf{Dataset: }  
A private longitudinal dataset was created from the Fellow Eyes of a real-world retrospective cohort of OCT scans from the PINNACLE consortium  \cite{sutton2022developing} collected from the University Hospital Southampton and Moorﬁelds Eye Hospital. The images were acquired using Topcon scanners 
with an average $3.6$ months interval between successive visits. A subset of the dataset was manually labelled for the TTC task and the remaining were used for training Morph-SSL.

The \textit{SSL Dataset} had 3570 unlabelled OCT scans from multiple visits of 399 eyes with at least 3 visits per eye. Whenever treatment information was available, the visits after the first anti-VEGF injection were removed to ensure that most scans in the dataset are in the iAMD stage. 

The \textit{TTC Dataset} with 343 Eyes (2418 OCT Volumes) was manually examined by clinical experts for the downstream task. In our experiments, each OCT scan was considered independently and the corresponding GT labels for $T^+$ (and $T^-$) were obtained as the time-interval between the current visit and the manually identified first visit of conversion (and the visit just before it). All Scans after the first visit of conversion were removed to focus on the iAMD stage and the earliest indicators of nAMD in the first conversion visit. 

\textbf{Preprocessing: }
The top and bottom boundaries delineating the retinal tissue called the Inner Limiting Membrane (ILM) and the Bruch's membrane (BM) were extracted using the automated method in \cite{garvin2009automated}. Thereafter, the curvature of the retinal surface was flattened by shifting each A-scan by an offset such that the BM lies on a straight plane similar to \cite{garvin2009automated}. The binary ROI mask of the retina contained the region from $26$ $\mu m$  above the ILM to $169$ $\mu m$ (to include the choroid) below the BM. Both the OCT and its ROI mask were then cropped to the central $3 \mytimes 3$ $mm^2$ en-face region. 
This region has been correlated with the onset of GA and neovascularization \cite{vogl2021spatio}. Finally, the volume was resized to $192 \mytimes 192 \mytimes 32$ and its intensity linearly scaled to [-1,1].

As an additional preprocessing for the \textit{SSL Dataset}, the enface projections of all visits of an eye were aligned to its first visit using the unsupervised affine registration method in \cite{vogl2021spatio}. This step ensures that the Morph-SSL features capture the structural changes caused by AMD progression instead of image misalignment. 
The step is not performed for the \textit{TTC Dataset} where each visit's scan is considered independently.

\begin{figure*}[!tbh]
 \centering
  \includegraphics[width=.8\textwidth]{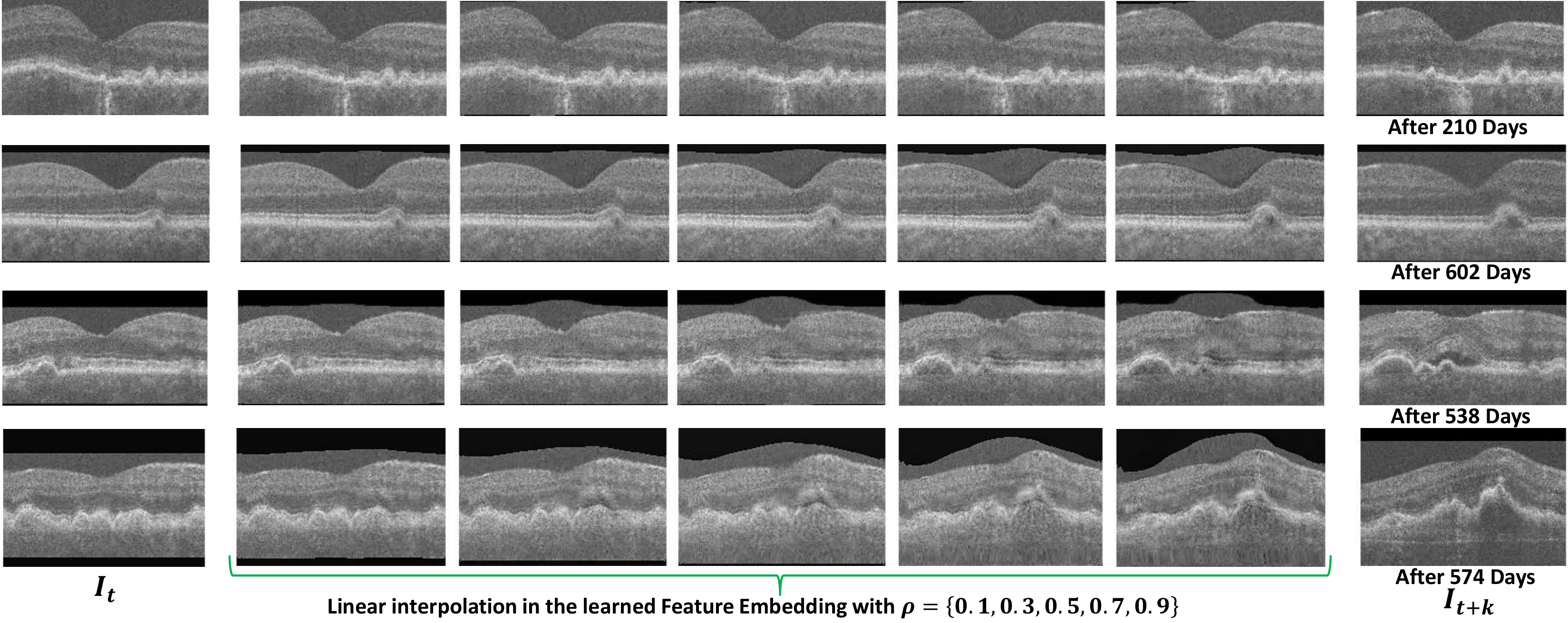}
\caption{\small Qualitative visualization of the linear interpolation between the features extracted from two OCT volumes $\mtrx{I}_t$ and $\mtrx{I}_{t+k}$ of the same eye. A single B-scan from the 3D volume has been depicted for a different eye in each row. The smooth transition in the generated intermediate images demonstrates Morph-SSL's ability to learn semantically meaningful features.}
\label{fig:interpolate} 
 \end{figure*}

\textbf{Experimental Setup: }
Morph-SSL was trained on image pairs formed from two random visits of the same eye, acquired within two years from each other. The \textit{SSL Dataset} was randomly divided into 350 eyes (14078 image pairs) for training, 25 eyes (640 image pairs) for validation and the remaining 24 eyes (600 image pairs) for a qualitative evaluation of the learned features (see Fig. \ref{fig:interpolate}).  

A stratified five-fold evaluation was conducted for the TTC task to reduce the bias of a specific train-test data split. The \textit{TTC Dataset} was randomly divided into 5 mutually exclusive parts at the eye level. The experiments were repeated 5 times, each time considering one part as the held out test set while the remaining dataset was randomly divided into $85\%$ for training and $15\%$ for validation. The performance was evaluated for predicting the conversion to nAMD within $t=0,6,12$ and $18$ months, where t=0 indicates that the input image is the first visit of conversion.
Area under the receiver operating characteristic curve (AUC) was used to evaluate the prediction scores and balanced accuracy computed as (Sensitivity+Specificity)/2, was used to evaluate the binary predictions. A single threshold was selected which maximized the average Youden's J statistic across $t=0,6,12,18$ months.

\textbf{Implementation Details: }
 All experiments were implemented in Python 3.8.5 with Pytorch 1.8.1 on a server, using a single NVIDIA A100, 40 GB GPU. The proposed Encoder and Decoder (including both Decoder-D and Decoder-A) had 3,757,616 and 907,721 network parameters respectively.  We kept the Decoder architecture small to force the Encoder to do most of the work in solving the Morph-SSL pretext task.
 
Both the Morph-SSL and downstream TTC training employed similar Data Augmentation comprising 
random 3D translations (up to $15\%$ of the image size along each axis), random horizontal flip (with $0.5$ probability), Gaussian blurring ($\mu\!=\!0$, random $\sigma \!\in\! [0,0.9]$) and Gaussian noise ($\mu\!=\!0$, $\sigma\!=\!0.001$). For Morph-SSL, both scans in the training image pair were translated and flipped identically, while other augmentations were applied independently.

During both training stages, Adam optimizer \cite{adam} was used ($\beta_1\!=\!0.9$, $\beta_2\!=\!0.999$, weight decay $\!=\!10^{-5}$ for Morph-SSL, $10^{-2}$ for TTC) with a cyclic learning rate schedule\cite{smith2017cyclical} where the learning rate was linearly varied from $lr_{min}$ ($10^{-6}$ for Morph-SSL, $10^{-5}$ for TTC) to $lr_{max}\!=\!10^{-4}$ and back to $lr_{min}$ in each epoch. We monitored validation performance at each epoch and saved the best network weights with minimum loss for Morph-SSL and highest average AUC for TTC. 

Morph-SSL trained with a batch size of 1 for 160 epochs, 2000 batch updates per epoch, required 23 GB GPU memory. The downstream training was performed for 400 epochs of 500 batch updates. A batch size of 6 was employed when the Encoder and Classifier were fine-tuned together on the TTC task, requiring 28 GB GPU memory. Training the Classifier alone required 4GB of GPU for a batch size of 16. 

\subsection{Results on the TTC task} 
\textbf{Impact of Morph-SSL: } In Table \ref{Tab:Result_Pred1}, rows 1-3, we evaluate 3 training setups: (a) end-to-end training from random  weight initialization; (b) freeze the Morph-SSL trained Encoder weights and only train the classifier on the TTC task; (c) use the Morph-SSL trained Encoder weights and the learned classifier weights from (b) to initialize and perform end-to-end finetuning of the Encoder and Classifier on the TTC task.

\begin{figure}[!tbh]
 \centering
  \includegraphics[width=.48\textwidth]{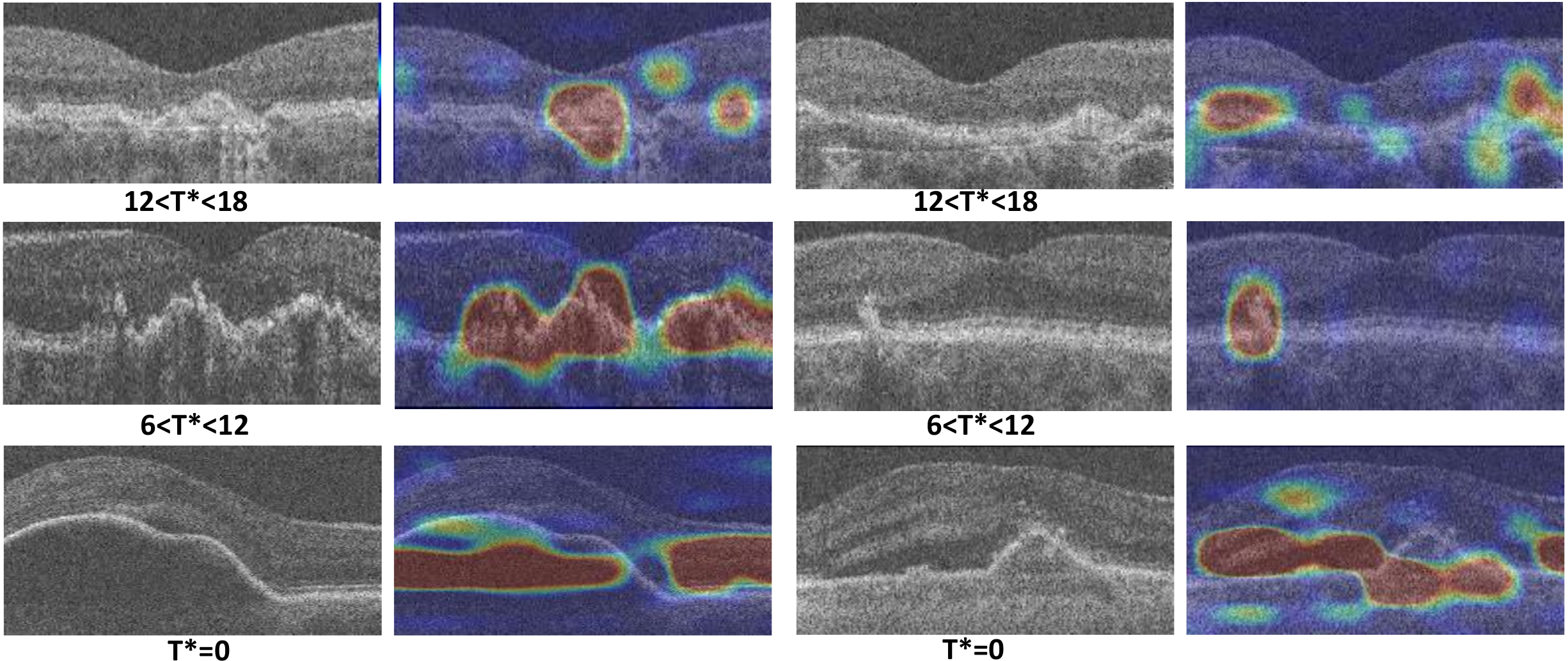}
\caption{\small Examples of Saliency maps with frozen Morph-SSL weights.}
\label{fig:sal} 
 \end{figure}

The Morph-SSL features showed significant performance improvement, even without fine-tuning, over end-to-end training form scratch (row 2 vs 1). Further fine-tuning on the TTC task (row 2 vs. 3) did not lead to a statistically significant improvement in AUC, except for $t=18$. This indicates that the initial Morph-SSL trained weights are very close to the optimal network weights for the TTC task. Overall, a good performance is observed in identifying the scans that have just converted to nAMD (t=0) or are about to convert within 6 months. However, the performance drops progressively as we consider larger time-intervals in the future. This may indicate that often, distinct morphological changes signaling imminent nAMD conversion appear unexpectedly only a few months before conversion rather than gradually over a long period. Few examples of the Saliency Maps $\mtrx{M}$ are shown in Fig. \ref{fig:sal}. 

\setlength{\tabcolsep}{5pt}
\renewcommand{\arraystretch}{1.2}
\begin{table*}[!tbh]
\centering
\caption{\small Area under the ROC curve (mean/std. deviation) across five-folds on the downstream task dataset. We test SSL methods under different training configurations by training on one-third or the entire dataset; freezing SSL-trained weights or finetuning end-to-end. DeLong test is performed between each state-of-the-art vs. the proposed method under identical training configurations, the values highlighted with $^*$ are \textbf{not} statistically different ($p>0.05$). The best performance in each column is highlighted in \textbf{bold}.}
\resizebox{0.985\textwidth}{!}{
\begin{tabular}{@{}l|llll|llll@{}}
\toprule
              & \multicolumn{4}{c|}{One-third Training data} & \multicolumn{4}{c}{Entire Training data} \\ 
              & \multicolumn{1}{c}{0 month} & \multicolumn{1}{c}{6 month} & \multicolumn{1}{c}{12 month} & \multicolumn{1}{c|}{18 month} & \multicolumn{1}{c}{0 month} & \multicolumn{1}{c}{6 month} & \multicolumn{1}{c}{12 month} & \multicolumn{1}{c}{18 month} \\ \midrule
Proposed-Freeze & $0.847\pm0.04$ & $0.742\pm0.04$ & $0.692\pm0.05$ & $0.672\pm0.06$               & $\mathbf{0.881\pm0.02}$ & $\mathbf{0.766\pm0.02}$ & $0.714\pm0.04$ & $0.687\pm0.06$\\

Proposed-Finetune & $\mathbf{0.869\pm0.02}$ & $\mathbf{0.763\pm0.03}$ & $\mathbf{0.710\pm0.04}$ & $\mathbf{0.682\pm0.05}$         & $0.876\pm0.02$ & $\mathbf{0.766\pm0.02}$ & $\mathbf{0.716\pm0.04}$ & $\mathbf{0.693\pm0.06}$ \\ \midrule

Model Genesis-Freeze \cite{model_genesis}& $0.806\pm0.03$ & $0.700\pm0.02$ & $0.655\pm0.01$ & $0.640\pm0.02$ & $0.825\pm0.02$ & $0.711\pm0.02$ & $0.664\pm0.02$ & $0.647\pm0.04$\\ 

Model Genesis-Finetune \cite{model_genesis}& $0.791\pm0.02$ & $0.698\pm0.04$ & $0.651\pm0.06$ & $0.623\pm0.06$ &    $0.850\pm0.03$ & $0.763\pm0.02^*$ & $0.709\pm0.04^*$ &  $0.690\pm0.06$ \\
 \midrule
 
Time prediction-Freeze \cite{rivail2019modeling}& $0.756\pm0.06$  & $0.664\pm0.03$ & $0.621\pm0.03$ & $0.588\pm0.03$                              & $0.775\pm0.02$ & $0.668\pm0.03$ & $0.619\pm0.02$   & $0.585\pm0.02$   \\
Time prediction-Finetune \cite{rivail2019modeling}& $0.789\pm0.03$  & $0.678\pm0.03$ & $0.621\pm0.01$ & $0.584\pm0.01$                                & $0.799\pm0.02$ & $0.683\pm0.03$ & $0.621\pm0.02$ & $0.581\pm0.02$   \\ \midrule

Barlow Twin-Freeze \cite{zbontar2021barlow}& $0.754\pm0.04$ &  $0.678\pm0.03$ &  $0.634\pm0.03$ & $0.609\pm0.04$              & $0.781\pm0.02$ & $0.670\pm0.02$ &  $0.602\pm0.03$  &  $0.562\pm0.05$ \\

Barlow Twin-Finetune \cite{zbontar2021barlow}& $0.780\pm0.03$ & $0.677\pm0.03$  & $0.640\pm0.04$  & $0.611\pm0.04$     & $0.780\pm0.04$ & $0.669\pm0.03$ & $0.628\pm0.05$ &  $0.604\pm0.06$ \\ \midrule

VICReg-Freeze \cite{bardes2021vicreg}& $0.788\pm0.04$ & $0.699\pm0.02$  & $0.665\pm0.03$  & $0.654\pm0.03$         & $0.846\pm0.02$ & $0.730\pm0.01$ &  $0.679\pm0.02$  & $0.651\pm0.03$  \\

VICReg-Finetune \cite{bardes2021vicreg}& $0.826\pm0.03$ & $0.737\pm0.02$  & $0.698\pm0.04$  & $0.679\pm0.04^*$  & $0.858\pm0.02^*$ & $0.754\pm0.01^*$ & $0.697\pm0.02$   & $0.675\pm0.03$  \\
 \bottomrule
\end{tabular}
}
\label{Tab:Result_SSL}
\end{table*}

\textbf{Impact of the loss terms for the TTC task: } An ablation of the auxiliary loss terms in  eq. \ref{eqn_ttc} is evaluated in rows 4, 5 of Table \ref{Tab:Result_Pred1}. Removal of the L-2 regularization on the slope parameter $a$ (row 2 vs 4) leads to a minor drop in the AUC across all 4 time-points. However, the effect on Balanced accuracy is mixed with slight performance drop at $t=6,12$ but improvement at $t=0,18$. Removing the loss term which penalizes high activations outside the retinal tissue leads to a small drop in both AUC and Balanced Accuracy for all time-points (row 5 vs 2).

\textbf{Impact of TTC formulation: }
We propose to model the CDF of the TTC with a sigmoidal function. An alternative way  is to pose it as multi-label classification with each class indicating if the image converts within a discrete time-point \cite{russakoff2019deep}, \cite{rivail2019modeling}, \cite{emre2022tinc}. We compare our performance against  multi-label classification in Table \ref{Tab:Result_Pred1}, row 6 by modifying the last layer of our classifier architecture to produce a 4 channel ouptut (instead of 1), to which GAP is applied followed by a sigmoid activation to obtain the predictions for the 4 time-points.

The AUCs of our method is slightly higher than multi-label classification at all time-points except $t\!=\!0$. The improvement is not statistically significant for $t=6,12$. However, our approach guarantees the monotonic non-decreasing property of the CDF (e.g., the probability of an eye to convert within 12 months cannot be lower than the conversion probability within 6 months) which is not the case with multi-label classification. Across the $5$ folds, the multi-label classifier is inconsistent in some cases, with higher predictions for a previous time-point compared to the next, $16$ cases between $t\!=\!0,6$ months, $60$ cases between $t\!=\!6,12$ and 84 cases with inconsistencies between $t\!=\!12,18$ months. Additionally, once trained, our model can predict conversion risk at any continuous time point within 18 months by varying $t$ in eq \ref{eqn:cdf}, unlike multi-label classification that can predict conversion risk only at predefined discrete time intervals used during training. 

\textbf{Architecture Design to predict slope: }
Spatial entropy of $\mtrx{M}$ was used to predict the slope $a$ of the sigmoid function. We compared this design choice against a modified architecture in Table \ref{Tab:Result_Pred1}, row 7 which predicts two channels. One channel is used similar to $\mtrx{M}$ to compute $b$ while a GAP is applied to the second channel for obtaining $a$. Although the new architecture requires extra network parameters, the difference in their AUCs (row 7 vs 2) was statistically insignificant.  

\textbf{Comparison with state-of-the art 3D Networks: }
An alternative to SSL is to fine-tune standard CNN networks after initializing them with the already available pre-trained weights. We compared our performance against two popular 3D-CNN networks, I3D \cite{carreira2017quo} and X3D \cite{feichtenhofer2020x3d}. Their last fully connected layer was modified to predict $a$ and $b$ in eq.\ref{eqn:cdf}. Both networks were initialized with pre-trained weights trained on the Kinetics video dataset and fine-tuned end-to-end on our task. Our Morph-SSL trained Encoder significantly outperformed both of these networks (row 2 vs 8,9 in Table \ref{Tab:Result_Pred1}) in terms of both AUC and Balanced accuracy across all time-points.

\textbf{Comparison with other SSL methods: }
We compare Morph-SSL against the state-of-the-art in Table \ref{Tab:Result_SSL}. The same 3D U-net and the transformations for the reconstruction task were employed for Model Genesis as reported in \cite{model_genesis}. The time interval prediction task \cite{rivail2019modeling} was originally developed for the central B-scans alone, however we implemented a 3D version using our Encoder architecture for a fair comparison. The latest CL methods, VICReg \cite{bardes2021vicreg} and Barlow Twins \cite{zbontar2021barlow} could not be trained in 3D due to their large batch size requirements. They were used to train a ResNet-50 with a batch size of 128 following \cite{emre2022tinc}. The positive image pairs were constructed by selecting B-scans (from the same position) from two random visits of the same eye within 18 months and applying the data augmentations used in \cite{emre2022tinc}. The Classifier was modified to handle a $2048\times 32$(feature dimensions $\mytimes$ B-scans) input. First, a 1D convolution layer with 32 input and 1 output feature channel was used to obtain a $2048$ dimensional feature for the entire OCT volume. This was followed by two fully connected layers with 1024 and 2 neurons respectively, to get the predictions for $a$ and $b$ in eq. \ref{eqn:cdf}. The SSL methods were compared under different training setups by: (a) using the SSL-trained features off-the-shelf and only training the Classifier (Freeze) vs. initialization with the SSL-trained network weights for end-to-end fine-tuning (Finetune), and (b) training on the entire vs. one-third of the supervised training data. To evaluate performance in a small data regime, one-third of the training data in each fold of the \textit{TTC Dataset} was randomly selected and kept consistent across all SSL methods. The Delong test was employed for statistical significance between AUCs using the pyroc 0.20 library \cite{pyroc}.

\noindent\textit{Small Data Regime:} Morph-SSL outperforms all benchmark methods (under identical Freeze/Finetune setup) across all 4 time-points in Table \ref{Tab:Result_SSL}. All differences were statistically significant except for VICReg-Finetune at $t\!=\!18$(p-value 0.30). Finetune improves performance over Freeze for all methods except Model-Genesis, which showed signs of overfitting. 

\textit{Entire Training Data:} In the Freeze setup, again Morph-SSL outperforms all benchmark methods in terms of AUC with a statistically significant difference across all 4 time-points. In the Fine-tune setup, Morph-SSL still clearly outperforms time-interval prediction \cite{rivail2019modeling} and Barlow twins \cite{zbontar2021barlow} with a statistically significant difference in AUC across all time-points. Although, Morph-SSL outperforms VICReg in terms of AUC the difference was not statistically significant for $t=0$ (p-value 0.27) and $t=6$ (p-value 0.28). Similarly, compared to Model-Genesis, the difference in AUC was not statistically significant for $t=6$ (p-value 0.21) and $t=12$ (p-value 0.41) with a marginally higher AUC for Morph-SSL. 

Overall, Morph-SSL shows better performance than other methods, particularly in scenarios where the features are used off-the-shelf or in a small data regime with limited labeled data for fine-tuning. When trained on the entire dataset, Morph-SSL was found to learn strong features with good performance on the TTC task with minimal effect of further fine-tuning.  

\subsection{Results on Risk Score for Progression to nAMD}
\begin{figure}[!h]
 \centering
  \includegraphics[width=0.35\textwidth]{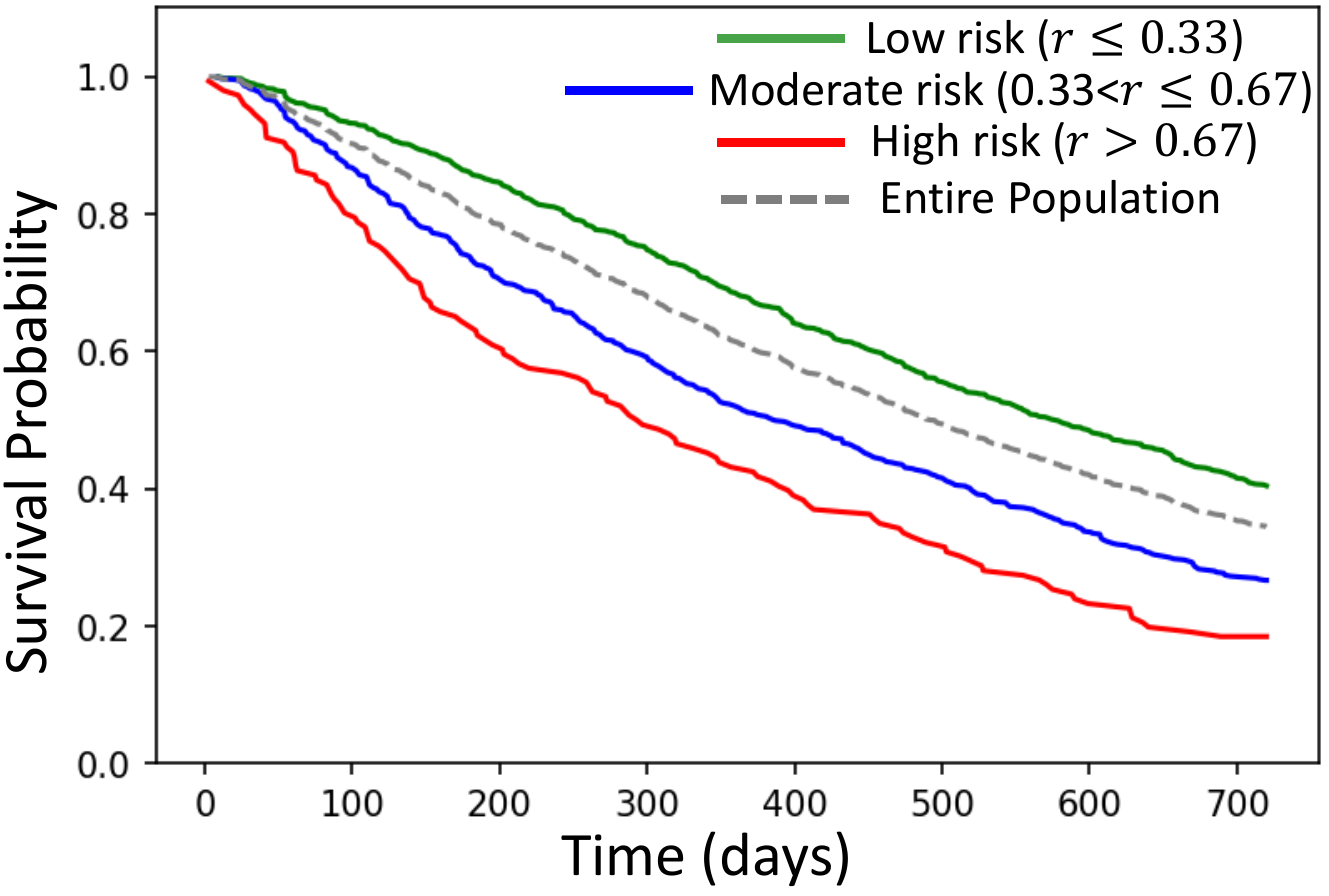}
\caption{Kaplan-Meier curves for different risk groups.}
\label{fig:kaplan_meier} 
 \end{figure}

A scalar risk score summarizing the risk of conversion to nAMD may be useful in clinical pratice. An ideal risk score should a) summarize the CDF of TTC into a single time independent scalar value; b) be bounded in the range $[0,1]$; c) be inversely proportional to the predicted time to conversion $b$. We formulate such a risk score by modifying eq. \ref{eqn:cdf} as $r=2/\left [1+exp \left \lbrace \frac{b}{a+0.05} \right \rbrace \right ]$. 
The test predictions for $a$ and $b$ were obtained from the five folds to compute $r$ for each OCT scan. The scans were then stratified into 3 groups with low risk ($0\le r\le0.33$), moderate risk ($0.33<r \le 0.67$) and high risk ($0.67 <r \le 1$). A population-level survival function for these groups is plotted in Fig. \ref{fig:kaplan_meier} using the Kaplan–Meier estimator on the GT conversion time. The survival curves for the three groups were found to be statistically well separated using the log-rank test with p value $<0.001$. Thus, $r$ is effective in stratifying eyes coming from different risk groups.

\subsection{Interpolation in the Morph-SSL feature space}
 Given a pair of scans $\mtrx{I}_{t}$, $\mtrx{I}_{t+k}$ from two visits of the same eye, we extract their features $\mtrx{F}_{t}$ and $\mtrx{F}_{t+k}$, and generate an intermediate feature through linear interpolation as $\mtrx{F}^{'}_{\rho}= \mtrx{F}_t + \rho. (\mtrx{F}_{t+1}-\mtrx{F}_{t})$, where $\rho \in [0,1]$. By using $\mtrx{F}_{t}$ and $\mtrx{F}^{'}_{\rho}$ (instead of $\mtrx{F}_{t+k}$ ) as inputs to the Morph-SSL trained Decoder, we can predict the transformation that morphs $\mtrx{I}_{t}$ to artificially generate the intermediate OCT scan for $\mtrx{F}^{'}_{\rho}$ (see Fig. \ref{fig:intro} c). The qualitative results in Fig. \ref{fig:interpolate} depict five intermediate scans by varying $\rho$. A gradual smooth transition between $\mtrx{I}_{t}$ and $\mtrx{I}_{t+k}$ is observed with the generated scans. Such a smooth feature embedding is enforced by our Decoder architecture which explicitly correlates the direction of the feature displacement $\mtrx{F}^{'}_{\rho}-\mtrx{F}_{t}$ to the \textit{type}, and its magnitude to the \textit{amount} of the morphing transformation. The magnitude increases with  $\rho$ while the direction remains the same.

This property may be explored in the future for different applications. Balanced-Mixup \cite{galdran2021balanced} generates artificial training samples by directly interpolating the voxels between two training images, which may produce blurry images. By interpolating in our feature embedding instead, more realistic samples may be generated. Another potential application could be to generate future OCT scans to visualize disease progression. A Recurrent Neural Network to predict the sequence of features of future visits may be explored for this task.
\section{Conclusion}
A vast amount of unlabelled longitudinal OCT scans are generated in clinics to monitor AMD. To leverage this data, we have proposed Morph-SSL, a novel SSL method designed to capture the temporal changes caused by disease progression. It ensures that the displacement in features between two OCT scans captures the morphological changes in the retina between them. The Encoder-Decoder network in Morph-SSL can be used to interpolate realistic intermediate scans between two visits. This offers promising future research directions for data augmentation and generating future OCT scans to visualize the expected trajectory of AMD progression. With the Morph-SSL trained Encoder, we have developed a prognostic model for TTC estimation that predicts the future risk of conversion from iAMD to nAMD from the current OCT scan. The lack of reliable biomarkers and wide variability in the rate of AMD progression makes it a challenging task. We modelled the CDF of TTC with a sigmoidal function over time. The Morph-SSL features were found to perform well on the TTC task even without fine-tuning and showed significant improvements over training from scratch or fine-tuning standard 3D-CNNs with pre-trained weights. It also outperformed popular SSL methods with significant gains in scenarios where SSL features are used off-the-shelf or fine-tuned on limited labeled data. Our method to predict the future risk of the onset of nAMD can play a critical role in enabling patient-specific disease management and enriching clinical trial populations with patients at risk.



\bibliographystyle{IEEE}
\bibliography{bibtex_small}

\end{document}